\documentclass{svproc}

\usepackage{url}

\usepackage{amsmath}
\usepackage{amssymb}
\usepackage{graphicx}
\usepackage{algorithm}
\usepackage[noend]{algpseudocode}
\usepackage{listings}
\def\mean#1{\left\langle #1 \right\rangle}

\def\tauint{\tau_\text{int}}
\def\CC{\hat C^{(c)}}

\begin{document}

\mainmatter              

\title{Everything you wish to know about time correlations but are
  afraid to ask}
\titlerunning{Correlation functions}  
%
\toctitle{Everything you wish to know about correlations but are
  afraid to ask}

\author{Tom\'as S.~Grigera}
\authorrunning{T.~S.~Grigera} 


\institute{Instituto de F\'\i{}sica de L\'\i{}quidos y Sistemas
  Biol\'ogicos (IFLySiB), Universidad Nacional de La Plata and CCT
  CONICET La Plata, Consejo Nacional de Investigaciones
  Cient\'\i{}ficas y T\'ecnicas,  Calle 59 no. 789, B1900BTE La Plata, Argentina.\\
  \and
  Departamento de F\'\i{}sica, Facultad de Ciencias Exactas,
  Universidad Nacional de La Plata, Argentina\\
\email{tgrigera@iflysib.unlp.edu.ar}, WWW home page:
\texttt{http://iflysib14.iflysib.unlp.edu.ar/tomas}
}

\maketitle              

\begin{abstract}

  We discuss the various definitions of time correlation functions and
  how to estimate them from experimental or simulation data.  We start
  with the various definitions, both in real and in Fourier space, and
  explain how to extract from them a characteristic time scale.  We
  then study how to estimate the correlation functions, i.e.\ how to
  obtain a good approximation to them from a sample of data obtained
  experimentally.  Finally we discuss some practical details that
  arise in the actual computation of these estimates, and we describe
  some relevant algorithms.



\end{abstract}

\section{Introduction}

This chapter is about the definitions and practical computation of
time correlation functions, i.e.\ the mathematical tools that enable
us to find and study dynamical correlations in physical systems.  Why
do we care about correlations? Science is about understanding how
things work (or, more ambitiously, how \emph{nature} works
\cite{bak1996}).  The question of what ``understanding'' something
really means is not one we plan to answer or even discuss here, but
most of us would probably agree that it involves (possibly among other
things) knowledge of a causation mechanism: to know how the state of
the system at some time influences the behavior of the same system at
a later time.  Now of course correlation does not imply causality: if
$A$ and $B$ are correlated, it might be that $A$ causes $B$, or that
$B$ causes $A$, or that something else causes both $A$ and $B$. So
correlations do not (directly) tell us about cause and effect.
However, causality is not directly measurable, while correlations are.
Correlations do not provide us with a causality mechanism, but do
constrain the cause-effect relationships we might care to imagine: to
explain how a system works, you are free to come up with whatever
mechanism (theory) you wish, but if that mechanism does not produce
the kind of correlations actually observed, then it cannot be right.

Correlations play a major role in the study of systems of many
particles, where the prediction (and maybe even the observation) of
detailed particle trajectories is out of the question, due to the
large number of variables involved.  Instead, one takes a statistical
approach, predicting (macroscopic) temporal correlations rather than
microscopic trajectories.  Conversely, in a macroscopic experiment one
does not measure particle trajectories, but records quantities that
are the result of the collective instantaneous state of many
microscopic components.  The exact variations of the observed
quantities are due to factors beyond our control (we say they are
\emph{noisy}), but the overall trends and the correlations between the
quantities measured at successive times furnish meaningful information
about the system's dynamics.  Moreover, time correlations provide
information on the dynamics even in the absence of overall trends
(e.g.\ when a physical system is in thermodynamic equilibrium).  But
time correlations are also greatly useful for the study of systems
quite more complex than physical systems in thermodynamic equilbrium:
in particular, of special interest for this volume, biological
systems.

We will thus proceed to present the mathematical definitions of time
correlation functions and their Fourier transforms (\S
\ref{sec:defin-time-corr}), to discuss the definition and meaning of
the correlation time (\S \ref{sec:relax-time-corr}) and finally to
turn to the question of computing time correlations from experimental
data (\S \ref{sec:comp-time-corr}).

But before moving on, I would like to make two clarifications.  First,
this chapter is about questions you are afraid to ask, not because
they are so advanced that they touch very dark and well-kept secrets,
but because they are so basic that you are embarrassed to ask.  No
honest scientific question should be embarrassing, but you know.
Second, this is perhaps not \emph{everything} you want to know about
time correlations.  In particular, I have not included any material
regarding their theoretical calculation.  But I do discuss all (well,
most) of what you need to know to compute them from experimental or
simulation data, starting from their various possible definitions and
touching many practical and sometimes nasty details.  I apologize if
these clarifications are disappointing, but you must agree that
``almost all you need to know to be able to compute time correlations
from experimental data, some of which is so basic that you are
embarrassed to ask'' would make for considerably less catchy title.

\section{Definition of time correlation functions}

\label{sec:defin-time-corr}

\subsection{Correlation and covariance}
\label{sec:covar-corr}

Let $x$ and $y$ be two random variables and $p(x,y)$ their joint
probability density.  We use $\langle\ldots\rangle$ to represent the
appropriate averages, e.g.\ the mean of $x$ is
$\mean{x}=\int x \, p(x) dx$ (recall that the probability distribution
of $x$ can be obtained from the joint probability,
$p(x)=\int\!\!dy\,p(x,y)$, and in this context is called marginal
probability) and its variance is
$\text{Var}_x=\bigl\langle (x-\mean{x})^2\bigr\rangle$. Let us write
\begin{equation}
  \label{eq:corr}
  C_{xy} = \langle x y \rangle = \int x y \, p(x,y) dxdy.
\end{equation}
Eq.~\eqref{eq:corr} defines the \emph{correlation} of $x$ and
$y$. Their \emph{covariance} is defined as
\begin{equation}
  \label{eq:covariance}
  \text{Cov}_{x,y} = \Bigl\langle \bigl( x-\mean{x}\bigr)
  \bigl(y-\mean{y}\bigr) \Bigr\rangle =
  \langle x y \rangle - \mean{x}\mean{y}.
\end{equation}
A property of the covariance is that it is bounded by the product of
the standard deviations (provided of course that they exist, which
cannot always be taken for granted):
\begin{equation}
  \label{eq:cov-ineq}
  \text{Cov}^2_{x,y} \le \text{Var}_x \text{Var}_y.
\end{equation}

The \emph{Pearson correlation coefficient,} or Pearson correlation for
short, is
\begin{equation}
  r_{x,y} = \frac{\text{Cov}_{x,y}}{\sqrt{\text{Var}_x\text{Var}_y}}.
\end{equation}
From the inequality (\ref{eq:cov-ineq}) it follows that the Pearson
coefficient is bounded, $-1\le r_{x,y}\le1$.  It can be shown that the
equality holds only when the relationship between $x$ and $y$ is
linear \cite[\S 2.12]{Priestley1981}.

The variables are said to be \emph{uncorrelated} if their covariance
is null:
\begin{equation}
  \label{eq:uncorrelated-def}
  \text{Cov}_{x,y}=0 \quad \Longleftrightarrow \quad
  \mean{xy} = \mean{x}\mean{y} \qquad \text{(uncorrelated)}.
\end{equation}
Absence of correlation implies that the variance of the sum is the sum
of the variances, because
\begin{equation}
  \text{Var}_{x+y} = \text{Var}_x + \text{Var}_y + 2\text{Cov}_{x,y},
\end{equation}
but is weaker than \emph{independence:} independence means that
$p(x,y)=p(x)p(y)$.  In the particular case when the joint probability
$p(x,y)$ is Gaussian, $\text{Cov}_{x,y}=0$ is equivalent to $x$ and
$y$ being independent, but this not true in general.  On the other
hand it is clear that independence does imply absence of correlation.
The covariance, or the correlation coefficient, can be said to measure
the degree of \emph{linear} association between $x$ and $y$, since it
is possible to build a nonlinear dependence of $x$ on $y$ that yields
zero covariance (see Ch.~2 of \cite{Priestley1981}).

\subsection{Fluctuating quantities as stochastic processes}
\label{sec:stochastic}

Consider now an experimentally observable quantity, $A$, that provides
some useful information on a property of a system of interest.  We
will usually assume here for simplicity that $A$ is a scalar, but the
present considerations can be rather easily generalized to vector
quantities.  $A$ can represent the magnetization of a material, the
number of bacteria in a certain culture, the rainfall in a certain
area, the prize of a share in the stock market, etc.  We assume that
$A$ can be measured repeatedly and locally in time, so that
we actually deal with a function $A(t)$.

We wish to compare the values of $A$ measured at different times.
However, we are interested in cases where $A$ is \emph{noisy,} i.e.\
subject to random fluctuations that arise because of our incomplete
knowledge of the variables affecting the system's evolution, or
because of our inability control them with enough precision (e.g.\ we
do not know all variables that can affect the variation of the price
of a stock market asset, we do not know all the interactions in a
magnetic system, we cannot control all the degrees of freedom of a
thermal bath that fixes the temperature of a sample).  The quantity
$A(t)$ is then a random variable, and to compare it at different
values of its argument we will resort to the correlation and
covariance defined in the previous section.  This will lead us to
define \emph{time correlation functions,} and this section is devoted
to stating their precise definitions.

Let us stress that a statement like ``$A(t_1)$ is larger than
$A(t_2)$'' is useless in practice, because even if it is meaningful
for a particular realization of the measurement process (experiment),
the noisy character of the observable means that a different
repetition of the experiment, under the same conditions, will yield a
different function $A(t)$.  Actually repeating the experiment may or
may not be feasible depending on the case, but we assume that we know
enough about the system to be able to assert that a hypothetical
repetition of the experiment would not exactly reproduce the original
$A(t)$.  For clarity, it may be easier to imagine that several copies
of the system are made and let evolve in parallel under the same
conditions, each copy then producing a signal slightly different from
that of the other copies.

We note that the expression ``under the same conditions'' is
implicitly qualified in some system-dependent way.  Clearly we expect
that two strictly identical copies of a system evolving under exactly
identical conditions will produce the same function $A(t)$.  Same
conditions here must be understood in a statistical way: the system is
prepared by choosing a random initial configuration extracted from a
well-defined probability distribution, or two identical copies evolve
with a random dynamics with known statistical properties (e.g.\
coupled to a thermal bath at given temperature and pressure).  We are
excluding from consideration cases where the fluctuations are mainly
due to experimental errors.  If that where the only source of noise,
one could in principle repeat the measurement enough times so that the
average $\langle A(t)\rangle$ is known with enough precision.
$\langle A(t)\rangle$ would then be an accurate description of the
system's actual evolution, and the correlations we are about to study
would be dominated by properties of the measurement process rather
than by the dynamics of the system itself.  Instead we are interested
in the opposite case: experimental error is negligible, and the
fluctuations of the observable are due to some process intrinsic to
the system.  Indeed in many cases (such as macroscopic observables of
systems in thermodynamic equilibrium) the average of the signal is
uninteresting (it's a constant), but the time correlation function
unveils interesting details of the system's dynamics.

\subsubsection{Stochastic processes}
\label{sec:stochastic-processes}

From our discussion of $A$ as a fluctuating quantity, it is clear that
$A(t)$ is not an ordinary function.  Rather, at each particular value
of time, $A(t)$ is a random variable, and $A(t)$ as a whole is a
random function, or \emph{stochastic process}.  A stochastic process
$A(t)$ is then a family of random variables indexed by $t$.  Thus
there must exist a family of probability densities $p(A,t)$ that
allows to compute all moments $\mean{A^n(t)}$ and in general any
average $\mean{f\bigl(A(t)\bigr)}$. However, $P(A,t)$ is not enough to
characterize the stochastic process, because in general the variables
$A(t)$ at different times are not independent.  Thus $p(A,t)$ is
actually a marginal distribution of some more complicated multivariate
probability density.

It is natural to imagine a functional $P\bigl[A(t)\bigr]$ that gives
the joint probability for all random variables $A(t)$.  However, an
infinite-dimensional probability distribution is a wild beast to ride,
and we shall content ourselves with the (infinite) set of $n$-variable
joint distributions
\begin{equation}
  \label{eq:4}
  P_n(A_1,t_1,A_2,t_2,\ldots,A_n,t_n).
\end{equation}
Most stochastic processes can be completely specified by the set of
all joint probabilities of the form (\ref{eq:4}) for all $n$ and all
possible choices of $t_1,\ldots,t_n$.  The sense of ``most'' is highly
technical \cite{Priestley1981}, but should include all processes of
interest to physicists.  Here we shall need only the set corresponding
to $n=2$ (which trivially gives also the set for $n=1$ marginalizing
on the second variable).

The notion of \emph{stationary processes} is an important one,
connected with the thermodynamic idea of equilibrium.
\emph{Completely stationary processes} are those for which all the
joint probability distributions that define the process are
translation invariant, that is
\begin{equation}
  P_n(A_1,t_1,\ldots,A_n,t_n) = P(A_1,t_1+s,\ldots,A_n,t_n+s),
  \qquad \forall s, n, t_i.
  \label{eq:5}
\end{equation}
A less restrictive notion is that of stationary processes \emph{up order
  $M$.}  This is defined by the requirement that all the joint moments
up to order $M$ exist and are time-translation invariant:
\begin{equation}
  \langle A^{m_1}(t_1)A^{m_2}(t_2)\ldots A^{m_n}(t_n)\rangle =
  \langle A^{m_1}(t_1+s)A^{m_2}(t_2+s)\ldots A^{m_n}(t_n+s)\rangle
\end{equation}
for all $s$, $n$, $\{t_1,\ldots,t_n\}$ and $\{m_1,\ldots,m_n\}$ such
that $m_1+m_2+\ldots+m_n\leq M$.  This is less restrictive not only
because of the bound on the number of random variables considered, but
also because invariance is imposed only on the moments, and not on the
joint distributions themselves.

In completely stationary processes the time origin is irrelevant: no
matter when one performs an experiment, the statistical properties of
the observed signal are always the same.  In particular, this implies
that the average $\langle A(t) \rangle$ is constant in time (but does
\emph{not} mean that time correlations are trivial).  This is the
situation one expects when observing a system in thermodynamic
equilibrium.  For processes stationary up to order $M$, time origin is
irrelevant for moments up to order $M$.  For a stationary process up
to first order, one can assert that $\langle A(t)\rangle$ is a
constant independent of $t$, and for a process stationary up to second
order one can in addition assert that $\langle A^2(t)\rangle$ (and
hence the variance) is time-independent, and that
$\langle A(t_1)A(t_2)\rangle = \langle A(0)A(t_2-t_1)\rangle$ (using
$s=-t_1$), i.e.\ that all second-order moments are function only of
the time difference.

\subsection{Time correlation functions}
\label{sec:time-correlations}

The time correlation function is the correlation of the random
variables $A(t_0)$ and $A(t_0+t)$, i.e.\ the second order moment
\begin{equation}
  \label{eq:def-timecorr}
C(t_0,t) = \langle A^*(t_0) A(t_0+t) \rangle  = \int\!\!dA_1\,dA_2\, 
 P(A_1,t_0,A_2,t_0+t) A_1^* A_2,
\end{equation}
where the star stands for complex conjugate. The time difference $t$
is sometimes called \emph{lag}, and the function is called \emph{self
  correlation} or \emph{autocorrelation} to emphasize the fact that it
is the correlation of the same observable quantity measured at
different times.  Note however that from the point of view of
probability theory $A(t_0)$ and $A(t_0+t)$ are two \emph{different}
(though usually not independent) random variables.  One can also
define \emph{cross-correlations} of two observables at different
times:
\begin{equation}
  \label{eq:6}
  C_{AB}(t_0,t) = \langle A^*(t_0) B(t_0+t) \rangle.
\end{equation}
The star again indicates complex conjugate.  From now on however we
shall restrict ourselves to real quantities and omit it in the
following equations.

One can also define a correlation function of the \emph{fluctations}
$\delta A(t)=A(t)-\mean{A(t)}$.  This is called the \emph{connected}
time correlation function\footnote{The name comes from diagrammatic
  perturbation theory, because it can be shown that only connected
  Feynman diagrams appear in the expansion of this quantity (see
  e.g.~\cite[ch.~8]{Binney1992}).},
\begin{equation}
  \label{eq:def-connected}
 C_c(t_0,t) = \mean{ \delta A(t_0) \delta A(t_0+t) } = C(t_0,t) -
 \mean{A(t_0)} \mean{A(t_0+t)}.
\end{equation}
Remembering (\ref{eq:uncorrelated-def}) it is clear that this function
is zero when the variables are uncorrelated (which we expect to happen
for $t\to\infty$).  For this reason this function is often more useful
than the correlation (\ref{eq:def-timecorr}), which for uncorrelated
variables takes the value $\mean{A(t_0)}\mean{A(t_0+t)}$.

The names employed here are usual in the physics literature
(e.g.~\cite{Binney1992,Hansen2013}).  In the mathematical statistics
literature, the connected correlation function is called
\emph{autocovariance} (in fact it is just the covariance of
$A(t_0)$ and $A(t_0+t)$), while the name autocorrelation is reserved
for the normalized connected correlation defined below
(\ref{eq:autocorr-math-stat}).

In what follows, unless otherwise stated, we will assume that we are
dealing with processes stationary at least up to second order, so that
the time correlation is a function only of the lag $t$.  In this case
the difference between the connected and nonconnected correlation
functions is a constant,
\begin{equation}
C_c(t) = \langle A(0)A(t)\rangle - \langle A\rangle^2 =  C(t) -
\langle A \rangle^2, \qquad \text{(stationary process).}
\label{eq:timecorr-stationary}
\end{equation}

Finally, let us define a normalized connected correlation so that its
absolute value is bounded by 1, in analogy with the Pearson coefficient:
\begin{align}
 \rho(t_0,t) &= \frac{C_c(t_0,t)}{\sqrt{C_c(t_0,0) C_c(t_0+t,0) }}, \\
 \rho(t) &= \frac{C_c(t)}{C_c(0)}, & \text{(stationary).}  
  \label{eq:autocorr-math-stat}
\end{align}

\subsubsection{Properties of the time correlation function}
\label{sec:prop-time-corr}

From the definition of the correlation function, it is easy to see
that in the stationary case the following hold:
\begin{enumerate}
\item $C_c(0)=\text{Var}_A$.
\item $|C_c(t)|\le C_c(0)$ for all $t$.
\item If $A(t)$ is real-valued, the correlation is even: $C(t)=C(-t)$.
\end{enumerate}

Also, for sufficiently long lags one expects that the variables become
independent and correlation is lost, so that
$C(t\to\infty)\to \langle A\rangle^2$.  Thus the connected correlation
should tend to zero, and it will typically have a Fourier transform
(see \S~\ref{sec:fourier-space})

\subsubsection{Example}
\label{sec:examples}

\begin{figure}
\includegraphics{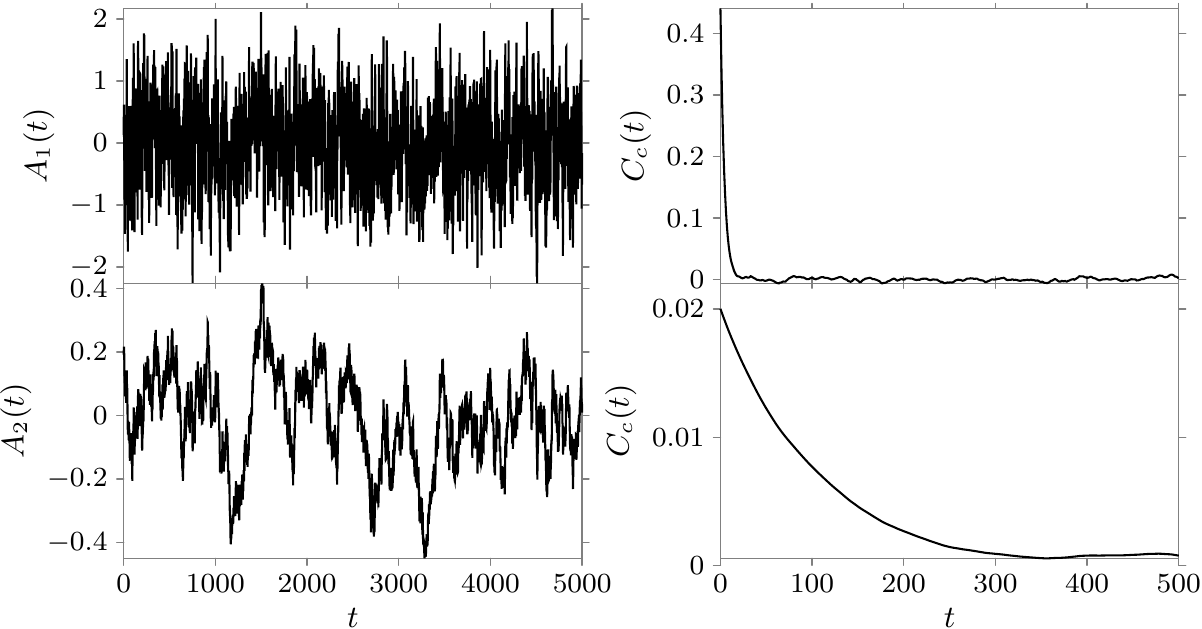}
\caption{Two stochastic signals (left) and their respective connected
  time correlations (right).  Correlation times are $\tau\approx 4.5$
  (top), $\tau\approx100$ (bottom).  The signals were generated
  through Eq.~\eqref{eq:test-sequence} with $\mu=0$, $\sigma^2=2$,
  $N=10^5$, and $w=0.8$ (top), $w=0.99$ (bottom).}
\label{fig:corr-example}
\end{figure}

Let us conclude this section of definitions with an example.
Fig.~\ref{fig:corr-example} shows on the left two synthetic
stochastic signals, generated with the random process described in \S
\ref{sec:example-synth-sign}.  On the right there are the
corresponding connected correlations.  The two signals look different,
and this difference is captured by $C_c(t)$.  We can say that
fluctuations for the lower signal are more persistent: when some value
of the signal is reached, it tends to stay at similar values for
longer, compared to the other signal.  This is reflected in the slower
decay of $C_c(t)$.

\subsection{Fourier transforms of time correlations}
\label{sec:fourier-space}

Time correlation functions are often studied in frequency space,
either because they are obtained experimentally in the frequency
domain (e.g.\ in techniques like dielectric spectroscopy), because the
Fourier transforms are easier to compute or handle analytically,
because they provide an easier or alternative interpretation of the
fluctuating process, or for other practical or theoretical reasons.
Although the substance is the same, the precise definitions used can
vary.  One must pay attention to i) the convention used for the
Fourier transform pairs and ii) the exact variables that are
transformed.

Here we define the Fourier transform pairs as
\begin{equation}
  \tilde f(\omega) = \int_{-\infty}^\infty\!\!dt\, f(t) e^{i\omega t}, \qquad
  f(t) = \int_{-\infty}^\infty\!\frac{d\omega}{2\pi}\,\tilde f(\omega) e^{- i\omega t},
\end{equation}
but some authors choose the opposite sign for the forward transform
and/or a different placement for the $1/2\pi$ factor (sometimes
splitting it between the forward and inverse transforms).  Depending
on the convention a factor $2\pi$ can appear or disappear in some
relations, like \eqref{eq:redu-almost-eq} below.

As for the second point above, time correlations are defined in terms
of two times, $t_1$ and $t_2$ (which we chose to write as $t_0$ and
$t_0+t$).  One can transform one or both of $t_1$ and $t_2$ or $t_0$
and $t$.  Let us consider two different choices, useful in different
circumstances.  First take the connected time correlation
\eqref{eq:def-connected} and do a Fourier transform on $t$:
\begin{equation}
  C_c(t_0,\omega) = \int\!\!dt\,e^{i\omega t} C_c(t_0,t_0+t) =
  e^{-i\omega t_0} \mean{\delta A(t_0) \delta \tilde A(\omega)},
\end{equation}
where $\delta \tilde A(\omega)$ stands for the Fourier transform of
$\delta A(t)$.  This definition is convenient when there is an
explicit dependence on $t_0$ but the evolution with $t_0$ is slow, as
in the physical aging of glassy systems (see
e.g.~\cite{cugliandolo_dynamics_2004}); one then studies a
time-dependent spectrum.  If on the other hand $C_c$ is stationary, it
is more convenient to write $t_1=t_0$, $t_2=t_0+t$ and do a double
transform in $t_1$, $t_2$:
\begin{multline}
  \tilde C_c(\omega_1,\omega_2) =  \int\!\!dt_1\,dt_2\,e^{i\omega_1
  t_1}e^{i\omega_2 t_2} C(t_1,t_2-t_1) = \\
  \int\!\!dt_1\,dt_2\,e^{i\omega_1 t_1}e^{i\omega_2 t_2} \mean{\delta
    A(t_1) \delta A(t_2)} = \mean{\delta\tilde A(\omega_1)\delta\tilde
    A(\omega_2)} =\\
  \int\!\!dt_1\,dt_2\,  e^{i(\omega_1+\omega_2) t_1} e^{i\omega_2 (t_2-t_1)} C_c(t_2-t_1) 
                          = (2\pi) \delta(\omega_1+\omega_2) \tilde C_c(\omega_2),
  \label{eq:fourier-transform-timecorr-double}
\end{multline}
where $\tilde C_c(\omega)$ is the Fourier transform of the stationary
connected correlation with respect to $t$ and we have used the
integral representation of Dirac's delta,
$\delta(\omega-\omega')=(2\pi)^{-1}\int_{-\infty}^\infty e^{i
  t(\omega-\omega')}\,dt$.  As the above shows, the transform is zero
unless $\omega_1=-\omega_2$.  For this reason it is useful to define
the \emph{reduced} correlation,
\begin{equation}
  C_c^R(\omega) = \mean{ \delta\tilde A(-\omega) \delta\tilde
                  A(\omega) } = \mean{ \delta\tilde A^*(\omega)
                  \delta\tilde A(\omega) },
\end{equation}
where the rightmost equality holds when $A$ is real.

The transform $\tilde C_c(\omega)$ is a well-defined function, because
the connected correlation decays to zero (and usually fast enough).
We can then say
\begin{equation}
  \label{eq:redu-almost-eq}
  \tilde C_c(\omega) = \frac{1}{2\pi\delta(0)} C_c^R(\omega).
\end{equation}
This relation furnishes a fast way to compute $C_c(t)$ in practice (\S
\ref{sec:algor-comp-estim}). But let us comment on the at first sight
baffling infinite factor relating the reduced correlation to
$\tilde C_c(\omega)$.  This originates in the fact that
$\delta\tilde A(\omega)$ cannot exist as an ordinary function, since
we have assumed that $A(t)$ is stationary.  This implies that the
signal has an infinite duration, and that, its statistical properties
being always the same, it cannot decay to zero for long times as
required for its Fourier transform to exist.  The Dirac delta can be
understood as originating from a limiting process where one considers
signals of duration $T$ (with suitable, usually periodic, boundary
conditions) and then takes $T\to\infty$.  Then
$2\pi \delta(\omega=0) = \int\!\!dt\,e^{i t \omega}|_{\omega=0} =
\int\!\!dt = T$.  These considerations can be made rigorous by
defining the signal's \emph{power spectrum,}
\begin{equation}
  \label{eq:power-spec}
  h(\omega)=\lim_{T\to\infty}\frac{1}{T}\mean{A_T(\omega)A_T^*(\omega)},
  \qquad A_T(\omega) = \int_{-T/2}^{T/2} A(t) e^{i\omega t} \,dt,
\end{equation}
and then proving that $h(\omega)=\tilde C_c(\omega)$ \cite[\S 4.7,
4.8]{Priestley1981}.

\section{Correlation time}
\label{sec:relax-time-corr}

The connected correlation function measures how correlation is
gradually lost as time elapses.  One often seeks for a summary of this
detailed information, in the form of a time scale that measures the
interval it takes for significant decorrelation to happen: this is the
\emph{correlation time}\footnote{The term relaxation time is often
  used interchangeably with correlation time.  The relaxation time is
  the timescale for the system to return to stationarity after an
  external perturbation is applied.  For systems in thermodynamic
  equilibrium, the fluctuation-dissipation theorem
  \cite{kubo_statistical_1998} implies that the two are equal.}
$\tau$.  While some definitions imply that after a correlation time the
connected correlation has descended to a prescribed level, this
quantity is usually better understood as a scale, which precise
meaning depends on the details of the shape of the correlation
function.  It is most useful to \emph{compare} correlation functions
of similar shape, which change as some environmental conditions varies
(e.g.\ when one studies a given system at a set of different
temperatures).  You should also keep in mind that correlation
functions can be quite complicated, and it may be appropriate to
describe them using more than one scale (e.g.\ they could be a
superposition of exponential decays).  Thus when one speaks of
\emph{the} correlation time one means (or should mean) the
\emph{largest} of them.  For the purpose of comparing different decays
to see which one is slower, it is less relevant which of the times one
chooses as description of the decay as long as a) one chooses the same
scale in all cases and b) one knows that all of them scale in the same
way when the control variable is changed.  We now discuss several
possible definitions of $\tau$.

\subsection{Correlation threshold}
\label{sec:threshold}

\begin{figure}
  \centering
  \includegraphics{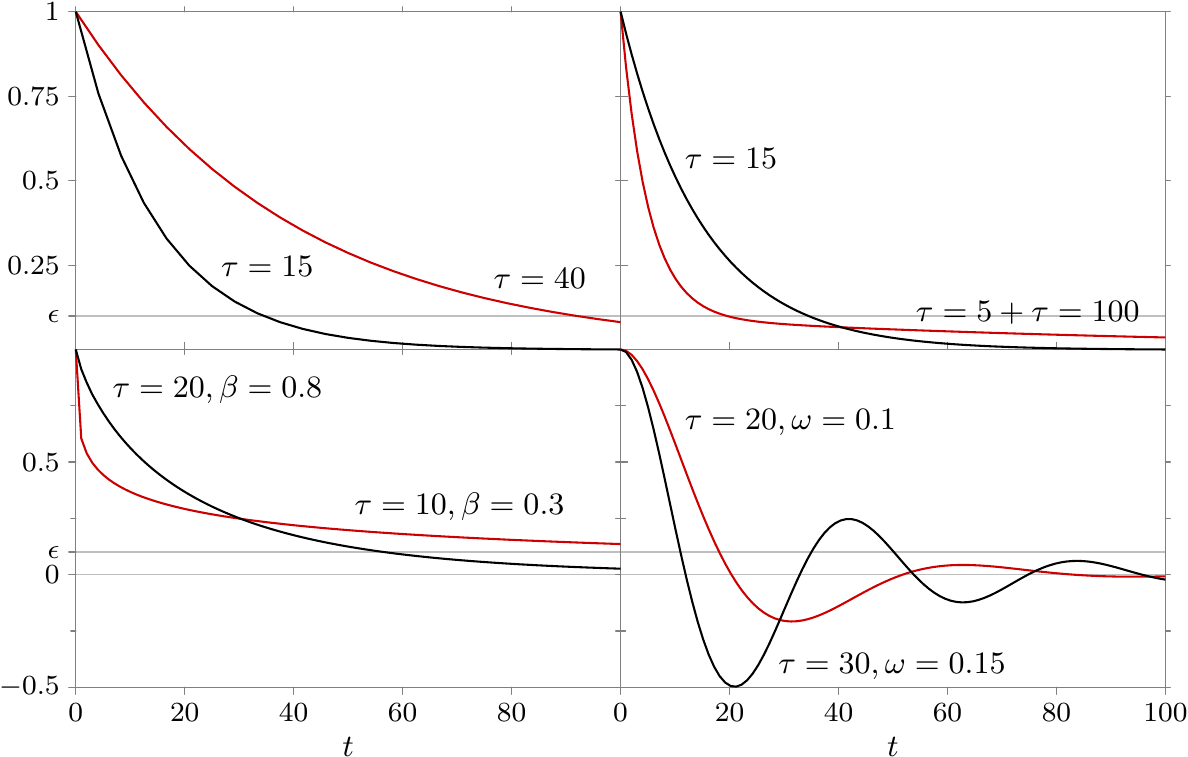}
  \caption{Different possible shapes for the decay of the time
    correlation. \textbf{Top left:} simple exponential.  \textbf{Top
      right:} simple exponential (black curve) and double exponential
    (red curve) decay.  In this case, the threshold criterion (here
    $\epsilon=0.1$) labels the red curve as the fastest, but it clearly
    has a longer tail.  \textbf{Bottom left:} stretched exponential.
    The average correlation time \eqref{eq:avetau} is
    $\mean{\tau}\approx22.7$ (black curve), $\mean{\tau}\approx92.6$
    (red curve).  \textbf{Bottom right:} exponentially damped
    harmonic oscillations. }
  \label{fig:shapes}
\end{figure}

A simple, ``quick and dirty'', way to define a correlation time is to
choose it as the time it takes for the (normalized, connected)
correlation to drop below a prescribed threshold $\epsilon$:
\begin{equation}
  \rho(t=\tau) = \epsilon.
\end{equation}
This definition is that is easy to apply to experimental data, and can
be useful to compare correlation functions as long as they have the
same shape.  But it can be ambiguous if the decay displays
oscillations, or if many time scales are present, and if it is
inadequate when applied to functions of different shapes (see
Fig.~\ref{fig:shapes}) or to power laws.

\subsection{Fit parameter}
\label{sec:fit-parameter}

Sometimes an analytical expression can be fit to the correlation
function, and a time scale extracted from the fit parameters. For
example, if one can fit an exponential decay,
\begin{equation}
   C_c(t)=A e^{-t/\tau},  \label{eq:simpl-exp}
\end{equation}
then the fitted value of the time scale $\tau$ can be used as
correlation time (which in this particular case coincides with the
threshold definition using $\epsilon=1/e$).  However, real-life time
correlations are usually more complicated than a simple exponential.
One can perhaps find more complicated functions that fit the
correlation, but be wary of the proliferation of
parameters\footnote{``With four parameters I can fit an elephant, and
  with five I can make him wiggle his trunk.'' Attributed to John von
  Neumann \cite{dyson2004}.} (as when fitting the sum of two, three,
$n$ exponentials).

You \emph{may} may get away with fitting the only the last part of
the decay with \eqref{eq:simpl-exp}, if you can fit a sizeable part of
the ``tail''.  The rationale is that the first part of the decay is
dominated by fast microscopic processes one is (often) not interested
in, so that by fitting the tail one obtains a good estimate of the
correlation time of the slowest process.  The problem with this
strategy is how to decide when the tail starts; if the $\tau$
obtained this way is too sensitive to how this choice is made then it
is probably no good.

A function widely used to describe non-exponential decays, employing
only three parameters, is the Kolrausch-William-Watts or
\emph{stretched exponential} function,
\begin{equation}
  C_c(t) = A e^{-(t/\tau)^\beta}.
\end{equation}
Here $\tau$ is a time scale and the stretching exponent $\beta$
controls the shape ($\beta<1$ gives a stretched exponential, i.e.\ a
function with a longer tail compared to a simple exponential of the
same $\tau$, while $\beta>1$ produces instead a compressed exponential).
However comparing the $\tau$ of two functions with different $\beta$
can be misleading (see Fig.~\ref{fig:shapes}).  A better description
of the decay using a single number can be achieved by considering the
correlation as a superposition of exponential processes,
\begin{equation}
  \rho(t) = \int_0^\infty w(\tau) e^{-t/\tau} \, d\tau,
\end{equation}
which defines the correlation time distribution function $w(\tau)$ as
essentially the inverse Laplace transform of $\rho(t)$
\cite{lindsey_detailed_1980}.  Then the average correlation time is
$\mean{\tau}=\int \tau w(\tau)\,d\tau$.  For the stretched exponential
one finds \cite{lindsey_detailed_1980}
\begin{equation}
  \mean{\tau}=\frac{\tau}{\beta}\Gamma\left(\frac{1}{\beta}\right),
  \label{eq:avetau}
\end{equation}
where $\Gamma(x)$ is Euler's gamma function.

\subsection{Integral time}
\label{sec:integral-time-sokal}

A quite general way to define a correlation time is from the integral
of the normalized connected correlation,
\begin{equation}
\tauint =  \int_0^\infty \rho(t) \, dt . \label{eq:tauint}
\end{equation}
Clearly for a pure exponential decay $\tauint=\tau$.  In general, if
$\rho(t)=f(t/\tau)$ then $\tauint = \text{const}\,\tau$.  The integral
time is related to the variance of the estimate of the mean of the
signal (see \S\ref{sec:estim-time-corr} below and
\cite[\S2]{sokal_monte_1997}).  With some care, it can be computed
from experimental or simulation data, avoiding the difficulties
encountered when using thresholds or fitting functions.  The procedure
explained next is expected to work well if long enough sequences are
at disposal and the decay does not display strong oscillations or
anticorrelation \cite{sokal_monte_1997}.

If $\CC_k\approx C_c(k\Delta t)$ is the estimate of the stationary
connected correlation (obtained as explained in
\S\ref{sec:estim-time-corr}), then the integral can be approximated by
a discrete sum, but the sum cannot run over all available values
$k=0,\ldots,N-1$, because as discussed in \S\ref{sec:estim-time-corr}
the variance of $\CC_k$ for $k$ near $N-1$ is large, so that the sum
$\sum_{k=0}^{N-1} \CC_k$ is dominated by statistical noise (more
precisely, its variance does not go to zero for $N\to\infty$).  A way
around this difficulty is \cite[\S3]{sokal_monte_1997} to cut-off the
integral at a finite time $t_c = c \Delta t$ such that $c\ll N$ but
the correlation is already small at $t=t_c$ (implying that $t_c$ is
larger than a few times $\tau$).  Thus $\tauint$ is defined
self-consistently as
\begin{equation}
\tauint =  \int_0^{\alpha \tauint} \rho(t) \, dt ,
\label{eq:tauauto}
\end{equation}
where $\alpha$ should be chosen larger than about $5$, and within a
region of values such that $\tauint(\alpha)$ is approximately
independent of $\alpha$.  Longer tails will require larger values of
$\alpha$; we have found adequate values to be as large as 20 in some
cases.  To solve \eqref{eq:tauauto} you can compute
$\tau(M)=\sum_k^M \CC_k/\CC_0$ starting with $M=1$ and increasing $M$
until $\alpha\tau(M)>M$.

\subsection{Correlation time from spectral content}
\label{sec:hh}

Another useful definition of correlation time can be obtained from the
Fourier representation of the normalized correlation,
\begin{equation}
  \tilde\rho(\omega)=\int_{-\infty}^\infty \!\!dt \rho(t) e^{i\omega
    t}.
\end{equation}
Because $\rho(t)$ is normalized,
$\int_{-\infty}^{\infty} \frac{d\omega}{2\pi} \,\tilde\rho(\omega)=1$,
a characteristic frequency $\omega_0$ (and a characteristic time
$\tau_0=1/\omega_0$) can be defined such that $\tilde\rho(\omega)$ for
$\omega\in[-\omega_0,\omega_0]$ holds half of the spectrum
\cite{halperin_scaling_1969}, i.e.\
\begin{equation}
 \int_{-\omega_0}^{\omega_0} \frac{d\omega}{2\pi} \,\tilde\rho(\omega)=\frac{1}{2}.
\end{equation}
This this definition of can be expressed directly in the time domain writing
\begin{equation}
 \int_{-\omega_0}^{\omega_0} \frac{d\omega}{2\pi}
 \int_{-\infty}^\infty \!\!dt \, \rho(t) e^{i\omega t} =
 2 \int_0^\infty \!\!dt \rho(t) \int_{-\omega_0}^{\omega_0} \frac{d\omega}{2\pi}
   e^{i\omega t} = \frac{2}{\pi} \int_{-\infty}^\infty \!\!dt \, \rho(t)
   \frac{\sin\omega_0 t}{t},
\end{equation}
where we have used the fact that $\rho(t)$ is even.  Then the
correlation time is defined by
\begin{equation}
  \label{eq:HHrelaxtime}
   \int_0^\infty \!\! \frac{dt}{t} \, \rho(t)
   \sin\left(\frac{t}{\tau_0}\right) = \frac{\pi}{4}.
\end{equation}
It can be seen that if $\rho(t) = f(t/\tau)$, then $\tau_0$ is
proportional to $\tau$ (it suffices to change the integration variable
to $u=t/\tau$ in the integral above).

An advantage of this definition is that it copes well with the case
when inertial effects are important and manifest in (damped)
oscillations of the correlation function (see Fig.~\ref{fig:shapes}).
In particular, for harmonic oscillations of frequency $\nu$,
$\tau_0=1/(2\pi\nu)$ while $\tauint$ is undefined.

\section{Computing time correlations from experimental data}

\label{sec:comp-time-corr}

In this section we examine in detail how to compute in practice the
time correlation function of a signal recorded in an experiment or
produced in numerical simulation.  Up till know we have discussed the
theoretical definitions of correlation functions, which are given in
terms of averages over some probability distribution, or ensemble.
However, when dealing with experimental or simulation data we do not
have direct access to the probability distribution, but only to a set
of samples, i.e.\ results from experiments, distributed randomly
according to an unknown distribution.  We must try to compute the
averages we want (the correlation functions), as accurately as
possible, using these samples.  This is what the field of statistics
is about: building \emph{estimators} that allow to compute the
quantities of interest as best as possible from the available samples.

We assume that the experiment records a scalar signal with a uniform
sampling interval $\Delta t$, so that we are handled $N$ real-valued
and time-ordered values forming a sequence $a_i$, with
\(i=1,\ldots, N\).  It is understood that if the data are digitally
sampled from a continuous time process, proper filtering has been
applied\footnote{According to the Nyquist sampling theorem, if the
  signal has frequency components higher than half the sampling
  frequency (i.e.\ if the Fourier transform is nonzero for
  $\omega\ge \pi /\Delta t$) then the signal cannot be accurately
  reconstructed from the discrete samples; in particular the high
  frequencies will ``polute'' the low frequencies (an effect called
  aliasing).  Thus the signal should be passed through an analog
  low-pass filter before sampling.  See \cite[\S 12.1]{Press1992a} for
  a concise self-contained discussion, or \cite[\S
  7.1]{Priestley1981}.}.  In what follows we shall measure time in
units of the sampling interval, so that in the formulas and algorithms
below we shall make no use of $\Delta t$.  To recover the original
time units one must simply remember that $a_i = A(t_i)$ with
$t_i=t_0 + (i-1) \Delta t$.  For the stationary case we shall write
$C_k=C(t_k)$ where $t_k$ is the time difference, $t_k= k \Delta t$ and
$k=0,\ldots,N-1$, and in the non-stationary case $C_{i,k}=C(t_i,t_k)$.

\subsection{Stationary vs.\ non-stationary signals}
\label{sec:stationary-vs.-non}

It is clearly hopeless to attempt to estimate an ensemble average
unless it is possible to obtain many samples under the same conditions
(i.e., if one is throwing dice, one should throw many times the
\emph{same} dice).  This implies there is a huge difference in how one
can treat a sample from a stationary process vs.\ a sample from a
non-stationary one.  If the process is stationary, we can essentially
consider the samples $a_i$ of one sequence as different repetitions of
the same experiment.  Estimation then basically consists in replacing
ensemble averages with time averages, e.g.\ one estimates the mean
$\mean{A}$ as $\overline{a}$, with
\begin{equation}
\overline{a}= \frac{1}{N} \sum_{i=0}^{N-1} a_i,
\label{eq:mean-estimate}
\end{equation}
and the stationary correlation with formula
\eqref{eq:conn-corr-estimate} below.  There is more to be said (see
\S \ref{sec:estim-time-corr} below), but in essence the situation
is this.

On the other hand, if the process is not stationary it means that the
samples of a single sequence are different experiments, in the sense
that the conditions have changed from one sample to another (i.e.\ the
system has evolved in some nontrivial way).  In this case the
correlation depends on two times (not just on their difference) and
the mean itself can depend on time.  The only way to estimate the mean
or the correlation function is to obtain many sequences $a_i^{(n)}$, $n=1,\ldots,M$
reinitializing the system to the same macroscopic conditions each time
(in a simulation, one can for example restart the simulation with the
same parameters but changing the random number seed).  Time averaging
is no good in this case, instead the ensemble average is replaced by
average across the different sequences, i.e.\ the (time-dependent)
mean is estimated as
\begin{equation}
  \mean{A(t_i)} \approx \overline{a_i} =\frac{1}{M} \sum_{n=1}^M a_i^{(n)},
\end{equation}
and the time correlation as
\begin{equation}
  C_c(t_i,t_k) \approx \CC_{i,k} =\frac{1}{M}\sum_n^M \delta a_i^{(n)}
  \delta a_{i+k}^{(n)}  , \qquad \delta a_i^{(n)} = a_i^{(n)} - \overline{a_i},
  \label{eq:conn-corr-ns-est}
\end{equation}
where the hat distinguishes an estimate from the actual quantity.
These estimators have the desirable property that they ``get better
and better'' when the number of samples $M$ grows, if the samples are
independent and the ensemble has finite variance.  More precisely,
$\CC_{i,k} \to C(t_i,t_i)$ as $M\to\infty$ (and similarly for
$\overline{a}$).  This property is called \emph{consistency,} (see
\cite[\S 5.2]{Priestley1981}) and is guaranteed because
$\mean{\CC_{i,k}}= C(t_i,t_i)$ and
$\mean{\overline{a_i}}=\mean{A(t_i)}$ (i.e.\ the estimators are
\emph{unbiased)} and their variance vanishes for $M\to\infty$ (because
the samples are independent).

If one has several sequences sampled from a stationary system, it is
possible to combine the two averages: one \emph{first} computes the
stationary correlation estimate \eqref{eq:conn-corr-estimate} for each
sequence, and \emph{then} averages the different correlation estimates
(over $n$ at fixed lag $k$).  It is clearly wrong to average the
sequences themselves before computing the correlation, as this will
tend to approach the (stationary) ensemble mean $\mean{A}$ for all
times, destroying the dynamical correlations.

Before turning to estimation in the stationary case, let us stress
that it is always sensible to check whether the sequence is actually
stationary.  A first check on the mean can be done dividing the
sequence in several sections and computing the average of each
section, then looking for a possible systematic trend.  If this check
passes, then one should compute the time correlation of each section
and then checking that all of them show the same decay (using fewer
sections than in the first check, as longer sections will be needed to
obtain meaningful estimates of the correlation function).  It is
important to note that this second check is necessary even if the
first one passes; as we have noted above it is possible for the mean
to be time-independent while the time correlation depends on two times
(i.e.\ the stochastic process is only stationary to first order).

\subsection{Estimation of the time correlation of a stationary process}
\label{sec:estim-time-corr}

If our samples $a_i$ are form a stationary signal, we build our
estimators using time averages in lieu of ensemble averages, 
in particular the estimator for $\mean{A}$ is
\eqref{eq:mean-estimate}.  As we said above, the idea is that we
regard the samples as different realizations of the same experiment.
However, they are \emph{not independent} realizations, but correlated
realizations.  The estimate \eqref{eq:mean-estimate} is still good in
the sense that it is consistent, but it has higher error than the
equivalent estimate built with independent samples, because it has a
higher variance.  The variance of the estimate with correlated samples
is \cite{sokal_monte_1997}
\begin{equation}
  \text{Var}_{\overline{a}} \approx \frac{2\tauint}{N} \left [
    \mean{A^2}-\mean{A}^2 \right],
\end{equation}
i.e.\ $2\tauint$ times larger than the variance in the independent
sample case, where $\tauint$ is the integral correlation time
\eqref{eq:tauint}.  In this sense $N/2\tauint$ can be thought of as
the number of ``effectively independent'' samples.

To estimate the connected correlation function we use
\begin{equation}
  \CC_k = \frac{1}{N-k} \sum_{j=1}^{N-k} \delta a_j \delta
a_{j+k}, \qquad \delta a_j = a_j - \overline{a}, \label{eq:conn-corr-estimate}
\end{equation}
where \(\overline{a}\) is usually the estimate
\eqref{eq:mean-estimate}, although the true sample mean $\mean{A}$ can
be used if known.  If the true mean is used, the estimator
\eqref{eq:conn-corr-estimate} is unbiased, otherwise it is
\emph{asymptotically} unbiased, i.e.\ the bias tends to zero for
$N\to\infty$, provided the Fourier transform of $C_c(t)$ exists.  More
precisely, $\langle \hat C_{c,k} \rangle = C_c(t_k) - \alpha/N$, where
$\alpha=2\pi \text{Var}_a \tilde C_c(\omega=0)$.  The \emph{variance}
of the estimate $\hat C_{c,k}$ is $O(1/(N-k))$ \cite[\S
5.3]{Priestley1981}.  This is sufficient to show that, at fixed $k$,
the estimator is consistent, i.e. $\CC_k\to C_c(t_k)$ for
$N\to\infty$.  However the variance increases for increasing $k$, and
thus the tail of $\hat C_{c,k}$ is considerably noisy.  In
practice, for $k$ near $N$ the estimate is mostly useless, and the
usual rule of thumb is to use $\hat C_{c,k}$ only for $k\le N/2$.

It is natural to propose an estimate for the nonconnected correlation
function doing a time average, analagous to \eqref{eq:conn-corr-estimate},
\begin{equation}
  \hat C_k  = \frac{1}{N-k} \sum_{j=1}^{N-k} a_j
             a_{j+k}.   \label{eq:corr-estimate} 
\end{equation}
However, although $\hat C_k$ is unbiased, it can have a large variance
when the signal has a mean value larger than the typical fluctuation
(see \S \ref{sec:two-prop-estim}).  In general, the connected
estimator should be used (an exception may be when an experiment can
only furnish many short independent samples of the signal, see the
example below in \S \ref{sec:example-synth-sign}).

Another asymptotically unbiased estimator of the connected correlation
can be obtained by using $1/N$ instead of $1/(N-i)$ as a the prefactor
of the sum in \eqref{eq:conn-corr-estimate}.  Calling this estimator
$C_{c,k}^*$, it holds that
$\langle C_{c,k}^*\rangle = C_c(t_k) - \alpha/N - k C(t_k)/N - \alpha
k/N^2$, where $\alpha$ is defined as before, and again $\alpha=0$ if
the exact sample mean is used.  This has the unpleasant property that
the bias depends on $k$, while the bias of $\hat C_{c,k}$ is
independent of its argument, and smaller in magnitude.  The advantage
of $C_{c,k}^*$ is that its variance is $O(1/N)$ independent of $k$,
thus it has a less noisy tail.  Many authors prefer $C_{c,k}^*$ due to
its smaller variance and to the fact that it strictly preserves
properties of the correlation (like property 2 of
\S~\ref{sec:prop-time-corr}), which may not hold for $\hat C_{c,k}$.
Here we stick to $\hat C_{c,k}$, as usual in the physics literature
(e.g.\ \cite{sokal_monte_1997,Allen1987,Newman1999}), so that we avoid
worrying about possible distortions of the shape at small $k$.  In
practice however, it seems that as long as $N$ is greater than
$\sim 10\tau$ (a necessary requirement in any case, see below), and
for $k\le N/2$, there is little numerical difference between the two
estimates.

\subsubsection{Two properties of the estimator}
\label{sec:two-prop-estim}

We must mention two important features of the estimator that are
highly relevant when attempting to compute numerically the time
correlation.  The first is that the difference between the
non-connected and connected estimators is not ${\overline a}^2$, but
\begin{equation}
 \hat C_k - \hat C_{c,k} = \overline{a} \left[ \frac{1}{N-k}
\sum_j^{N-k} a_j + \frac{1}{N-k}\sum_j^{N-k} a_{j+k}
-\overline{a} \right],
\end{equation}
as it is not difficult to compute.  The difference does tend to
\(\overline{a}^2\) for \(N\to\infty\), but in a finite sample
fluctuations are important, especially at large $k$.  Fluctuations are
additionally amplified by a factor $\overline a$, so that when the
signal mean is large with respect to its variance, the estimate
$\hat C_k$ is completely washed out by statistical fluctuations.  In
practice, this means that while it is true that $C(t)$ and $C_c(t)$
differ by a constant term ($\mean{A}^2$), in general it is a bad idea
to compute $\hat C_k$ and subtract $\overline{a}^2$ to obtain an
estimate of the connected correlation.  Instead, \emph{one computes
  the connected correlation directly} (by computing first an estimate
of the mean and using \eqref{eq:conn-corr-estimate}) and then, if
needed, adds $\overline{a}^2$ to obtain an estimation of $C(t)$.

Another important fact is that the estimator of the connected
correlation \emph{will always have zero,} whatever the length of the
sample used, \emph{even if $N\ll \tau$.}  To see this, consider the quantity
\begin{equation}
B_i = (N-i) \hat C_{c,i} =  \sum_{j=1}^{N-i} \delta a_j \delta a_{j+i},
\end{equation}
and compute the sum
\begin{equation}
\sum_{i=0}^{N-1} B_i = \sum_{i=0}^{N-1} \sum_{j=1}^{N-i} \delta a_j
\delta a_{j+i} = \sum_{i=0}^{N-1} \sum_{j=1}^{N} \sum_{k=1}^{N} \delta a_j
\delta a_k \delta_{k,j+i}.
\end{equation}
But \(\sum_{i=0}^{N-1} \delta_{k,j+i}\) equals 1 if \(k\ge j\) and 0, so
\begin{multline}
\sum_{i=0}^{N-1} B_i = \sum_{j=1}^N\sum_{k=j}^N
\delta a_j \delta a_k = \frac{1}{2} \sum_{j\neq k}^N \delta a_j \delta
a_k + \sum_{j=1}^N (\delta a_j)^2 = \\
\frac{1}{2} \left[ \sum_{j=1}^N
\delta a_j \right]^2 + \frac{1}{2}\sum_{j=1}^N (\delta a_j)^2 =
\frac{1}{2}\sum_{j=1}^N (\delta a_j)^2 > 0,
\end{multline}
where the last equality follows because \(\sum_j \delta a_j=0\).
Now we can easily do the same sum starting from $i=1$:
\begin{equation}
  \sum_{i=1}^{N-1} B_i = -B_0 + \sum_{i=0}^{N-1} B_i =
  - \frac{1}{2}\sum_{j=1}^N (\delta a_j)^2 <0.
\end{equation}
This shows that at least some of the $B_i$ must be negative.  But
since $B_0>0$, the conclusion is that $B_k$, and hence $\hat C_{c,k}$,
which differs from it by a positive factor, \emph{must change sign at
  least once for $k\ge1$.} 

The practical consequence of this is that when $N$ is of the order of
$\tau$, or smaller, the estimate $\hat C_{c,k}$ will suffer from
strong finite-size effects, and its shape will be quite different from
the actual $C_c(t)$.  In particular, since $\hat C_{c,k}$ will
intersect the $x$-axis, it can mislead you into thinking that $\tau$
is smaller than $N$ when in fact it is several times larger.  Be
suspicious if $\hat C_{c,k}$ changes sign once and stays very
anticorrelated.  Anticorrelation may be a feature of the actual
$C_c(t)$, but if the sample is long enough, the estimate should decay
until correlation is lost (noisily oscillating around zero).  One must
always perform tests estimating the correlation with different values
of $N$: if the shape of the correlation at short times depends on $N$,
then $N$ must be increased until one finds that estimates for
different values of $N$ coincide for lags up to a few times the
correlation time (see example below).

Once a sample-size-independent estimate has been obtained, the
correlation time can be estimated (\S \ref{sec:relax-time-corr}), and
it must be checked that self-consistently $N$ is several times larger
than $\tau$.

\subsubsection{Example}
\label{sec:example-synth-sign}

To illustrate the above considerations, we generate a correlated
sequence from the recursion
\begin{equation}
 a_i = w a_{i-1} + (1-w) \xi_i,  \label{eq:test-sequence}
\end{equation}
where $w\in[0,1]$ is a parameter and the $\xi_i$ are uncorrelated Gaussian
random variables with mean $\mu$ and variance $\sigma^2$.  Assuming
the signal is stationary it is not difficult to find
\begin{equation}
  \mean{a}=\mu, \qquad \mean{(a-\mu)^2}=\frac{1-w}{1+w}\sigma^2, \qquad C^{(c)}_k
  =\mean{(a_i-\mu)(a_{i+k}-\mu)} = \sigma^2 w^k,
\end{equation}
so that the correlation time is $\tau=-1/\log w$.

\begin{figure}
  \centering
  \includegraphics{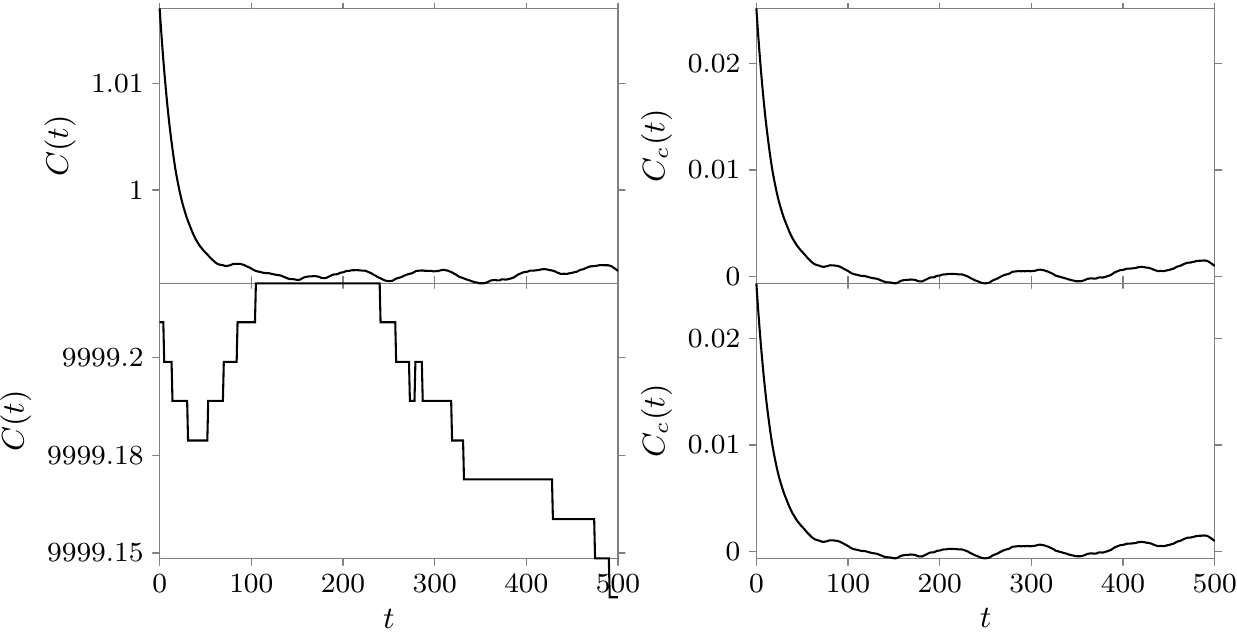}
  \caption{Connected vs.\ nonconnected estimate.  The estimate of
    $C(t)$ (equation \eqref{eq:corr-estimate}, left panels) is much
    worse that the estimates of $C_c(t)$ (equation
    \eqref{eq:conn-corr-estimate}, right panels).  The (artificial)
    signal was generated with \eqref{eq:test-sequence}. Top panels:
    $\mu=1$; bottom panels: $\mu=100$.  In both cases, $\sigma^2=1$,
    $w=0.95$ ($\tau\approx20$) and length $N=5\cdot 10^4$. }
  \label{fig:conn-vs-nonconn}
\end{figure}

We used the above recursion to generate artificial sequences and
computed their time correlation functions with the estimates discussed
above.  Fig.~\ref{fig:conn-vs-nonconn} shows the problem that can face
the non-connected estimate.  When the average of the signal is smaller
than or of the order of the noise amplitude (as in the top panels),
one can get away with using \eqref{eq:corr-estimate}.  However if
$\mu\gg\sigma$, the non-connected estimate is swamped by noise, while
the connected estimate is essentially unaffected (bottom panels).
Hence, if one is considering only one sequence, one should always use
the connected estimator.  The situation might be different if many
sequences, corresponding to different realizations of the same
experiment, are available (see the example after next).

\begin{figure}
  \centering
  \includegraphics{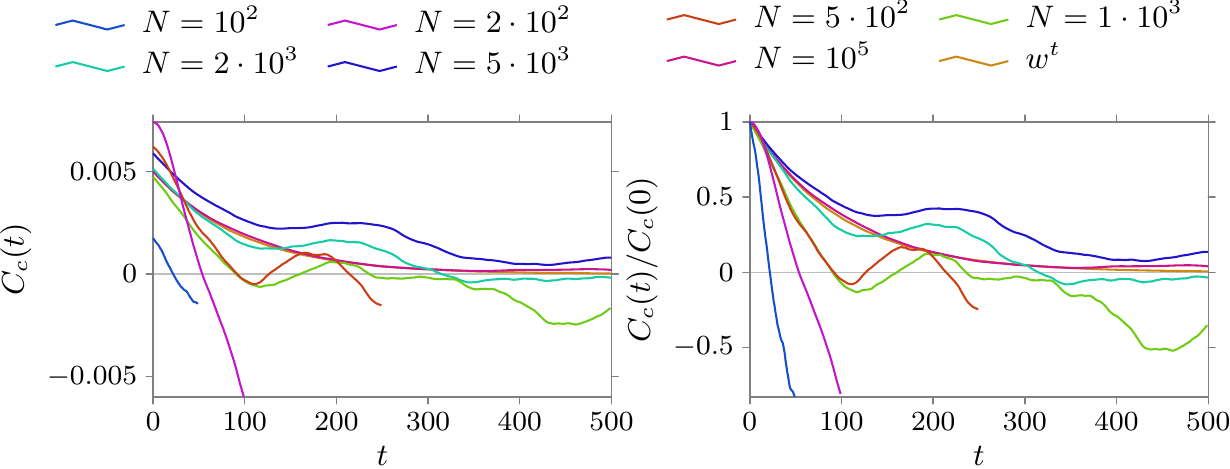}
  \caption{Finite size effects. Estimates of the connected correlation
    (top) and normalized connected correlation (bottom) for sequence
    \eqref{eq:test-sequence} of different lengths as indicated,
    together with the analytical result $C_c(t)=\sigma^2
    w^t$. Parameters are $\mu=0$, $\sigma^2=1$, $w=0.99$
    ($\tau\approx100$). }
  \label{fig:finite-size}
\end{figure}

In Fig.~\ref{fig:finite-size} we see how using samples that are too
short affects the correlation estimates.  The same artificial signal
was generated with different lengths.  For the shorter lengths, it is
seen that the correlation estimate crosses the $t$ axis (as we have
shown it must) but does not show signs of losing correlation.  One
might hope that $N=1000\approx 10\tau$ is enough (the estimate starts
to show a flat tail), but comparing to the result of doubling the
length shows that it is still suffering from finite-length effects. It
is seen that a sequence at least $20\tau$ to $50\tau$ long is needed
to get the initial part of the normalized connected correlation more
or less right, while a length of about 1000$\,\tau$ is necessary to
obtain a good estimate up to $t\sim5\tau$.  The unnormalized estimator
suffers still more from finite size due to the increased error in the
determination of the variance (left panel).

\begin{figure}
  \centering
  \includegraphics{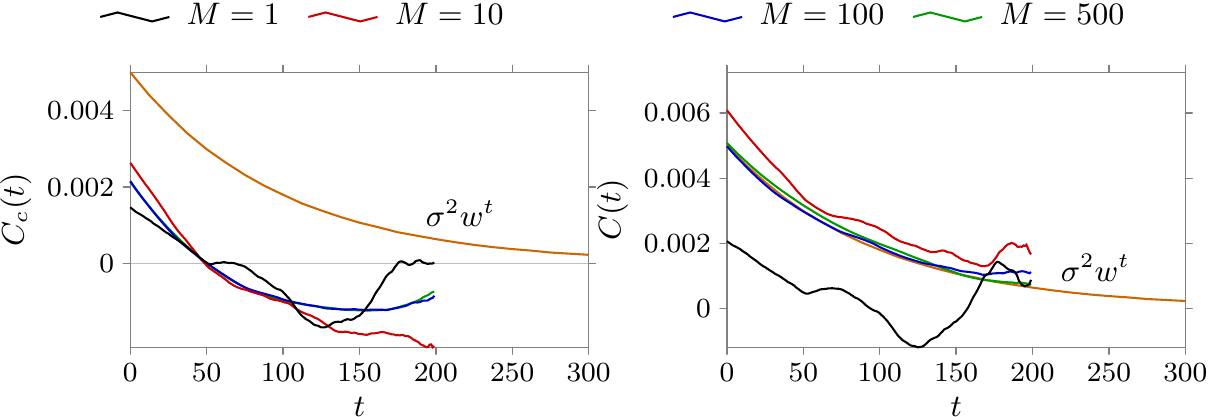}
  \caption{Effect of averaging the estimates of many short sequences.
    We show the connected (left) and non-connected (right) estimates
    of $M$ different samples of sequence \eqref{eq:test-sequence} as
    indicated in the legend, with $N=200$, $\mu=0$, $\sigma^2=1$,
    $w=0.99$ ($\tau\approx100$).}
  \label{fig:many-short}
\end{figure}

If the experimental situation is such that it is impossible to obtain
sequences much longer than the correlation time, one can get some
information on the time correlation if it is possible to repeat the
experiment so as to obtain several \emph{independent} and
\emph{statistically equivalent} sample sequences.  In
Fig.~\ref{fig:many-short} take several (say $M$) sequences with the
same parameters as in the previous example, but quite short
($N=200\approx 2\tau$).  As we have shown, it is not possible to
obtain a moderately reasonable estimate of $C_c(t)$ using one such
sequence (as is also clear from the $M=1$ case of
Fig.~\ref{fig:many-short}, where the analytical result is plotted for
comparison).  However, the figure shows how one may benefit from the
multiple repetitions of the (short) experiment by averaging together
all the estimates.  The averaged connected estimates are always far
from the actual correlation function, even for $M=500$ (the case which
contains in total $10^5$ points, which proved quite satisfactory in
the previous example): this is consequence of the fact that all
connected estimates must become negative.  Instead, the averaged
non-connected estimates approach quite closely the analytical
result even though not reaching the decorrelated region\footnote{Note
  that in this case $\mu=0$ so that fluctuations are larger than the
  average.  If that were not the case, one may attempt to compute a
  connected correlation estimate by using all sequences to estimate
  the average, then substracting this same average to all sequences.}.
Although it is tricky to try to extract a correlation time from this
estimate (one may fit the initial decay but it is not possible to know
whether the final decay will follow this trend or whether some very
slow tail is present), this procedure at least offers a way to obtain
some dynamical information in the face of experimental limitations.

\subsection{Algorithms to compute the estimators}
\label{sec:algor-comp-estim}

The estimators can be computed numerically by straightforward
implementation of equations \eqref{eq:conn-corr-ns-est}
or~\eqref{eq:conn-corr-estimate}, although in the stationary case it
is much more efficient to compute the connected correlation through
relation \eqref{eq:power-spec} using a fast Fourier transform (FFT)
algorithm.  Let us focus on the stationary case and examine in some
detail these algorithms.

Algorithm \ref{algo:direct} presents the direct method.  It is
straightforward to translate the pseudo-code to an actual language of
your choice.  Apart from some missing variable declarations, the only
thing to consider is that it is probably not convenient (or even
illegal in some languages, as in classic C) to return a large array,
and it is better to define $C$ as an output argument, using a pointer
or reference (as e.g.\ FORTRAN or C do by default) to avoid copying
large blocks of data.  The advantages of this algorithm are that it is
self-contained and simple to understand and implement.  Its main
disadvantage is that, due to the double loop of lines 8--11, it runs
in a time that grows as $N^2$.  For $N$ up to about $10^5$ this
algorithm is perfectly fine: a good implementation in a compiled
language should should run in a few seconds in a modern computer.  But
this time grows quickly; in the author's computer $N=5\cdot10^5$ takes
35 seconds, for $N=10^6$ the time is two and a half minutes.  In
contrast, the algorithm with FFT takes 1 second for $N=10^6$ and 11
seconds for $N=10^7$.

\begin{algorithm}
  \caption{Compute the connected correlation of sequence $a$ (of
    length $N$) using the direct $O(N^2)$ method.  The connected
    correlation is returned in vector $C$.}
  \label{algo:direct}
  \begin{algorithmic}[1]
    \Function{timecorr}{$a$,$N$}
    \State $\mu \gets 0$   \Comment{Compute average}
    \For {$i=1,\ldots,N$} \State $\mu \gets \mu+a_i$ \EndFor
    \State $\mu \gets \mu/N$
    \For {$i=1,\ldots,N$} \Comment{Clear $C$ vector}
    \State $C_i \gets 0$ \EndFor

    \Statex

    \For {$i=1,\ldots,N$} \Comment{Correlation loop} \State
    $d \gets a_i-\mu$ \For {$k=0,\ldots,N-i$} \State
    $C_{k+1} \gets C_{k+1} + d * (a_{i+k}- \mu)$ \EndFor \EndFor

    \Statex
   
    \For {$i=1,\ldots,N$} \Comment{Normalize and return} \State
    $C_i \gets C_i / (N-i-1)$ \EndFor \State \textbf{return} C
    \EndFunction
  \end{algorithmic}
\end{algorithm}

So, if you need the correlation of really long sequences, the
FFT-based algorithm, though more difficult to get running, pays off
with huge savings in CPU time at essentially the same numerical
precision.  The idea of the algorithm is to compute the Fourier
transform of the signal, use \eqref{eq:redu-almost-eq} to obtain the
Fourier transform of the connected correlation, then transform back to
obtain $C_c(t)$.  This is faster than algorithm~\ref{algo:direct}
because the clever FFT algorithm can compute the Fourier transform in
a time that is $O(N\log N)$.

Actually, we need discrete versions of the Fourier transform formulas
(as we remarked before, the Fourier transform of the continuous time
signal does not exist).  The \emph{discrete Fourier transform (DFT)} and its
inverse operation are defined \cite[\S 12.1]{Press1992a} as (it is
convenient to let the subindex of $a_i$ run from $0$ to $N-1$ to write
the following two equations),
\begin{equation}
  \tilde a_k = \sum_{j=0}^{N-1} e^{2\pi i j k/N} a_j, \qquad
  a_j =\frac{1}{N} \sum_{k=0}^{N-1} e^{- 2\pi i j k/N} \tilde a_k,
\end{equation}
where we note that the inverse DFT effectively extends the sequence
periodically (with period $N$).  The discrete version of
\eqref{eq:redu-almost-eq} is \cite[\S 13.2]{Press1992a}
\begin{align}
  \tilde D_k & = | \tilde a_k |^2, & \text{where} &&
  D_j & = \sum_{k=0}^{N-1} a_k a_{k+j},
\end{align}
and where the definition of $D_j$ makes use of the (assumed)
periodicity of $a_i$.  $D_j$ is almost our estimate
\eqref{eq:conn-corr-estimate}: we only need to take care of the
normalization and of the fact that due to the assumed periodicity of
$a_i$ some past times are regarded as future, e.g.\ for $k=10$, in the
sum there appear the terms $a_0a_{10}$ up to $a_{N-11}a_{N-1}$ (which
are fine), but also $a_{N-10}a_0$ through $a_{N-1}a_9$, which we do
not want included.  This is fixed by padding the original signal with
$N$ zeros at the end, i.e.\ setting $a_k=0$ for $k=N,\ldots,2N-1$ and
ignoring the values of $D_j$ for $j\ge N$.

In summary, to compute the connected correlation using FFT the steps
are i) estimate the mean and substract from the $a_i$, ii) add $N$
zeros to the end of the sequence, iii) compute the DFT of the
sequence, iv) compute the squared modulus of the transform, iv)
compute the inverse DFT of the squared modulus, v) multiply by the
$1/(N-i)$ prefactor.  Pseudocode for this algorithm is presented as
algorithm~\ref{algo:FFT}.

\begin{algorithm}
  \caption{Compute the connected correlation of sequence $a$ (of
    length $N$) using a fast Fourier transform.  This algorithm is
     $O(N\log N)$.}
  \label{algo:FFT}
  \begin{algorithmic}[1]
    \Function{timecorr}{$a$,$N$} 
    \State $\mu \gets 0$ \Comment{Compute average}
    \For {$i=1,\ldots,N$}
    \State $\mu \gets \mu+a_i$ \EndFor
    \State  $\mu \gets \mu/N$
    \For {$i=0,\ldots,N$} \Comment{Substract the average from signal}
    \State $a_i \gets a_i-\mu$
    \EndFor
    \For {$i=N,\ldots,2N$}\Comment{Pad with 0s at the end}
    \State $\text{a}_i \gets 0 $
    \EndFor
    \Statex

    \State $b\gets$\Call{FFT}{$a$,$2N$} \Comment{Compute the FFT of a
      as a vector of length $2N$} \For {$i=1,\ldots,2N$}
    \Comment{Compute squared modulus of $b$} \State
    $b_i\gets \lvert b_i\rvert^2$ \Comment{Note that the Fourier
      transform is complex} \EndFor \State
    $C \gets$\Call{IFFT}{$b$,$2N$} \Comment{Inverse FFT} \Statex
 
    \State $C\gets$\Call{resize}{$C$,$N$} \Comment{Discard the last
      $N$ elements of $C$} \For {$i=1,\ldots,N$} \Comment{Normalize
      and return}
    \State $\text{C}_i \gets \text{C}_i / (N-i-1)$ \EndFor \State
    \textbf{return} C \EndFunction
  \end{algorithmic}
\end{algorithm}

To translate this into an actual programming language the comments
made for algorithm~\ref{algo:direct} apply, and in addition some extra
work is needed for lines 10--13.  First, one needs to choose an FFT
routine.  If you are curious about the FFT algorithm, you can read for
example \cite[Ch.~12]{Press1992a} or \cite{duhamel1990}, but writing
an FFT routine is not easy, and implementing a state-of-the-art FFT is
stuff for professionals.  Excellent free-software implementations of
the FFT can be found on Internet.  FFTW \cite{frigo2005}, at
\texttt{http://www.fftw.org} deserves mention as particularly
efficient, although it is a large library and a bit complex to use.
Note that some simpler implementations require that $N$ be a power of
two, failing or using a slow $O(N^2)$ algorithm if the requiriment is
not fulfilled.  Also pay attention to i) the difference between
``inverse'' and ``backward'' DFTs (the latter lacks the $1/N$ factor),
ii) how the routine expects the data to be placed in the input array,
iii) how it is returned, and iv) whether the transform is done
``in place'' (i.e.\ overwriting the original data) or not.  If the
routine is a ``complex FFT'' it will expect complex input data (so
that for real sequences you will have to set to zero the imaginary
part of the $a_i$), while if it is a ``real FFT'' routine it will
typically arrange (``pack'') the output data in some way, making use
of the discrete equivalent of the $\tilde A(-\omega)=A^*(\omega)$
symmetry so as to return $N$ real numbers instead of $2N$ (the real
and imaginary parts of the complete DFT).  This affects the way you
must compute the squared modulus (lines 11--12).  For example, for the
packing used by the FFTW real routines, lines 11--12 translate to (in
C)
\begin{lstlisting}
  b[0]*=b[0];
  for (int i=1; i<N; i++) {
      b[i] = b[i]*b[i] + b[2*N-i]*b[2*N-i];
      b[2*N-i] = 0;
  }
  b[N]*=b[N];  
\end{lstlisting}

\section{Conclusion}

I have tried to convey the basic notions about time correlation
functions as well as some practical advice on how to compute them from
actual data.  I hope this account will be useful for students and
researchers finding themselves in need to compute time correlations.

On closing, I wish to thank the colleagues with whom I have
worked over the years, and which have helped shape my understanding of
time correlations through discussions on concepts and practicalities,
in particular A.~Cavagna, I.~Giardina, V.~Mart\'\i{}n-Mayor, G.~Parisi
and the late P.~Verrocchio.

Last but not least, I thank D.~Chialvo for contributing the title, for
organizing and inviting me to the Complexity Weekend, for agreeing
that a tutorial of this sort should be written, and for editing and
getting this volume to the press.

\bibliography{main}
\bibliographystyle{spphys}

\end{document}